\documentstyle[aps,epsf,preprint]{revtex}
\tightenlines
\begin{document}
\draft
\title{Properties of the light scalar mesons face the experimental data on the $\phi\to\pi^0\pi^0\gamma$ decay and the $\pi\pi$ scattering}
\author
{N.N. Achasov
\thanks{achasov@math.nsc.ru}
\ and A.V. Kiselev
\thanks{kiselev@math.nsc.ru}}

\address{
   Laboratory of Theoretical Physics,
 Sobolev Institute for Mathematics, Novosibirsk, 630090}

\date{\today}
\maketitle

\begin{abstract}
The high-statistical KLOE data on the $\phi\to\pi^0\pi^0\gamma$
decay are described simultaneously with the data on the $\pi\pi$
scattering and the $\pi\pi\to K\bar K$ reaction. The description
is carried out taking into account the chiral shielding of the
putative $\sigma (600)$ meson and it's mixing with the
well-established $f_0(980)$ meson. It is shown that the data don't
contradict the existence of the $\sigma (600)$ meson and yield
evidence in favor of the four-quark nature of the $\sigma (600)$
and $f_0(980)$ mesons.
\end{abstract}

\section{Introduction}
Study of the nature of light scalar resonances is one of the
central problems of non-perturbative QCD. The point is that the
elucidation of their nature is important for understanding both
the confinement physics and the chiral symmetry realization way in
the low energy region, i.e., the main consequences of QCD in the
hadron world. Actually, what kind of interaction at low energy is
the result of the confinement in the chiral limit? Is QCD
equivalent to the non-linear $\sigma$-model or the linear one at
low energy?

The experimental nonet of the light scalar mesons Ref.
\cite{pdg-2004}, the putative $f_0(600)$ (or $\sigma (600)$) and
$\kappa (700-900)$ mesons and the well-established $f_0(980)$ and
$a_0(980)$ mesons,  suggests the $U_L(3)\times U_R(3)$ linear
$\sigma$-model  \cite{nature}. The history of linear
$\sigma$-model is rather long, so that the list of its
participants, quoted  in Ref. \cite{gellman}, is far from
complete.

Hunting the light $\sigma$ and $\kappa$ mesons had begun in the
sixties already, and a preliminary information on the light scalar
mesons in Particle Data Group Reviews had appeared at that time.
But long-standing unsuccessful attempts to prove their existence
in a conclusive way entailed general disappointment, and
information on these states disappeared from Particle Data Group
Reviews. One of the principal reasons against the $\sigma$ and
$\kappa$ mesons was the fact that both  $\pi\pi$ and $\pi\kappa$
scattering phase shifts do not pass over $90^0$ at putative
resonance masses. The situation changes when it was shown in Ref.
\cite{annshgn-94} that in the linear $\sigma$-model there is a
negative background phase which hides the $\sigma$ meson
\cite{kaminski}. It has been made clear that shielding of wide
lightest scalar mesons in chiral dynamics is very natural. This
idea was picked up (see, for example, Ref. \cite{ishida}) and
triggered a new wave of theoretical and experimental searches for
the $\sigma$ and $\kappa$ mesons, see Particle Data Group Review
\cite{pdg-2004}.

In theory the principal problem is an impossibility to use the
linear $\sigma$-model in the tree level approximation inserting
widths into $\sigma$ meson propagators Ref. \cite{annshgn-94}
because such an approach breaks  both the unitarity and the Adler
self-consistency conditions Ref. \cite{adler-65b}. Strictly
speaking, the comparison with the experiment requires the
non-perturbative calculation of the process amplitudes Ref.
\cite{approxim}.

Nevertheless, now there are the possibilities to estimate odds of
the $U_L(3)\times U_R(3)$ linear $\sigma$-model to the underlying
physics of light scalar mesons in phenomenology. Really, even now
there is a huge body of information about the $S$-waves of
different two-particle pseudoscalar states. What is more, the
relevant information goes to press almost continuously from BES,
BNL, CERN, CESR, DA$\Phi$NE, FNAL, KEK, SLAC and others. As for
theory, we know quite a lot about the scenario under discussion:
the nine scalar mesons, the putative chiral masking
\cite{annshgn-94} of the $\sigma (600)$ and $\kappa( 700-900)$
mesons, the unitarity, the analiticity and the Adler
self-consistency conditions. In addition, there is the light
scalar meson treatment motivated by field theory. The foundations
of this approach were formulated in
Refs.\cite{adsh-79,adsh-80,adsh-1980,z_phys,adsh-84}, see also
Refs. \cite{bog_conf,our_prop}, Ref. \cite{approach}. In
particular, the propagators of scalar mesons were introduced in
this approach. As was shown in Ref. \cite{our_prop}, these
propagators satisfy the K\"allen -- Lehmann representation in the
wide domain of coupling constants of the light scalar mesons with
two-particle states. The present paper is the first step in the
realization of this plan.

 The $a_0(980)$
and $f_0(980)$ scalar mesons, discovered more than thirty years
ago, became the hard problem for the naive quark-antiquark ($q\bar
q$) model from the outset. Really, on the one hand the almost
exact mass degeneration  of the isovector $a_0(980)$ and isoscalar
$f_0(980)$ states revealed seemingly the structure similar to the
structure of the vector $\rho$ and $\omega$ mesons, S-wave $q\bar
q$ states, or as $f_2(1270)$, $a_2(1320)$ and $f_1(1285)$,
$a_1(1260)$ - P-wave $q\bar q$ states, what is closer to our case
of scalar mesons. On the other hand, the coupling of $f_0(980)$
with the $K\bar K$ channel pointed unambiguously to a considerable
part of the strange quark pair $s\bar s$ in the wave function of
$f_0(980)$. It was noted in the late 1970s that in the MIT bag
model (which incorporates confinement phenomenologically) there
are light four-quark scalar states and it was suggested that
$a_0(980)$ and $f_0(980)$ might be these states \cite{jaffe},
containing the $s\bar s$ pair additionaly to the non-strange one.
>From that time $a_0(980)$ and $f_0(980)$ resonances became the
subject of  intensive investigations, see, for example, Refs.
\cite{achasov-2002,achasov-2003,dop2,ishida,adsh-79,adsh-80,adsh-1980,z_phys,adsh-84,bog_conf,our_prop,montanet,achasov-84,achasov-89,bramon,achasov-97,achasov-97a,achasov-97b,achasov-98,snd-1,snd-2,snd-fit,snd-ivan,cmd,publ,pi0publ,kloe,a0f0,achasov-01a,bramon-2002,achasov-95,mistake,lmas,our,our_a0,achsh-2004}.

Ten years later it was shown  \cite{achasov-89} that the study of
the  radiative decays $\phi\to a_0\gamma\to\eta\pi^0\gamma$ and
$\phi\to f_0\gamma\to\pi^0\pi^0\gamma$ can shed light on the
puzzle of the light scalar mesons. Over the next ten years before
the experiments (1998), this question was examined from different
points of view
\cite{bramon,achasov-97,achasov-97a,achasov-97b,achasov-98}.

Now these decays have been studied not only theoretically but also
experimentally. The first measurements have been reported by the
SND \cite{snd-1,snd-2,snd-fit,snd-ivan} and CMD-2 \cite{cmd}
Collaborations which obtain the following branching ratios
$$Br(\phi\to\gamma\pi^0\eta)=(8.8\pm 1.4\pm0.9)\times 10^{-5}\
\mbox{\cite{snd-fit}},$$
 $$Br(\phi\to\gamma\pi^0\pi^0)=
(12.21\pm0.98\pm0.61)\times 10^{-5}\ \mbox{\cite{snd-ivan}},$$
 $$Br(\phi\to\gamma\pi^0\eta)=(9.0\pm 2.4\pm 1.0)\times 10^{-5} \ \mbox{\cite{cmd}},$$
$$Br(\phi\to\gamma\pi^0\pi^0)=(9.2\pm0.8\pm0.6)\times 10^{-5}\
\mbox{\cite{cmd}}.$$

In Refs. \cite{snd-2,snd-ivan,cmd} the data on the
$\phi\to\pi^0\pi^0\gamma$ decay were analyzed in the $K^+K^-$ loop
model of the single $f_0 (980)$ resonance production, suggested in
Ref. \cite{achasov-89}. In Ref. \cite{a0f0} the data on the
$\phi\to\pi^0\pi^0\gamma$ decay and on the $\delta^0_0$ phase  of
the $\pi\pi$ scattering were analyzed simultaneously for the first
time, which allowed one to determine relative phase between the
signal amplitude $\phi\to S\gamma\to\pi^0\pi^0\gamma$ and the
background one $\phi\to \rho\pi^0\to\pi^0\pi^0\gamma$ properly.
The consideration was carried out in the model suggested in Ref.
\cite{achasov-97}. This model is based on the the $K^+K^-$ loop
model \cite{achasov-89} of the $\pi\pi$ production and involves
constructing  the $K\bar K\to\pi\pi$ amplitude both above and
under the $K\bar K$ threshold. Recall that the phase of the $K\bar
K\to\pi\pi$ amplitude under the $K\bar K$ threshold equals the
phase of the $\pi\pi$ scattering.

Then the data came from the KLOE experiment \cite{publ,pi0publ}:
$$ Br(\phi\to\gamma\pi^0\eta)=(8.51\pm 0.51\pm0.57)\times 10^{-5}\
\mbox{in}\  \eta\to\gamma\gamma\ \mbox{\cite{publ}},$$
$$Br(\phi\to\gamma\pi^0\eta)=(7.96\pm0.60\pm0.40)\times 10^{-5}\
\mbox{in}\ \eta\to\pi^+\pi^-\pi^0\ \mbox{\cite{publ}},$$
$$Br(\phi\to\gamma\pi^0\pi^0)= (10.9\pm0.3\pm 0.5)\times 10^{-5}\
\mbox{\cite{pi0publ}},$$

\noindent in agreement with the Novosibirsk data
\cite{snd-1,snd-2,snd-fit,snd-ivan,cmd} but with a considerably
smaller error. Note that the reanalysis of the KLOE data on the
$\phi\to\gamma\pi^0\eta$ can be found in Ref. \cite{our_a0}.

Unfortunately, in Ref. \cite{pi0publ} interference of the signal
reaction $e^+e^-\to\phi\to\pi^0\pi^0\gamma$ with the coherent
background $e^+e^-\to\omega\pi^0\to\pi^0\pi^0\gamma$ was not taken
into account. So the data in the region of low invariant $\pi\pi$
masses, $m_{\pi\pi}$,  are not correct even by the order of
magnitude  \cite{thanks}. The interference under discussion makes
separation of the signal reaction in the low-mass region difficult
in general. The data in the high-mass region ( $m>$660 MeV ) can
be treated as correct since they suffer no $\omega\pi^0$ effect.

In this paper we show that the KLOE data on the
$\phi\to\pi^0\pi^0\gamma$ decay and the data on the $\pi\pi$
scattering and the $\pi\pi\to K\bar K$ reaction up to 1.1 GeV can
be described in the upgraded model of Ref. \cite{achasov-97},
taking into account the chiral shielding of the $\sigma (600)$
meson and it's mixing with the $f_0(980)$ meson.

All formulas for the $\phi\to(S\gamma+\rho^0\pi^0)\to\pi^0
\pi^0\gamma$ reaction ($S=f_0(980)+\sigma(600)$)  are shown in
Sec.\ref{sf}. The results of the data analysis are presented in
Sec.\ref{sr}.  A brief summary is given in Sec.\ref{sc}.

\section{The formalism of the \lowercase{$\phi\to
(f_0(980)+\sigma(600))\gamma\to\gamma\pi^0\pi^0$} and
\lowercase{$\phi\to\rho^0\pi^0\to\gamma\pi^0\pi^0$} reactions
 }
\label{sf}

In Refs. \cite{achasov-89,achasov-97} it was shown that the
dominant background process is
$\phi\to\pi^0\rho\to\gamma\pi^0\pi^0$, while the reactions
$e^+e^-\to\rho\to\pi^0\omega\to\gamma\pi^0\pi^0$ and
$e^+e^-\to\omega\to\pi^0\rho\to\gamma\pi^0\pi^0$ have a small
effect on $e^+e^-\to\phi\to\gamma\pi^0\pi^0$ in the region
$m_{\pi^0\pi^0}\equiv m>900$ MeV. In Ref. \cite{a0f0} it was shown
that the $\phi\to\pi^0\rho\to\gamma\pi^0\pi^0$ background is small
in comparison with the signal $\phi\to\gamma
f_0(980)\to\gamma\pi^0\pi^0$ at $m>700$ MeV.

The amplitude of the background decay
$\phi(p)\to\pi^0\rho\to\gamma(q)\pi^0(k_1)\pi^0(k_2)$ has the
following form:

\begin{equation}
M_{back}=e^{-i\delta}g_{\rho\pi^0\phi}g_{\rho\pi^0\gamma}\phi_{\alpha}p_{\nu}
\epsilon_{\delta}q_{\epsilon}\epsilon_{\alpha\beta\mu\nu}
\epsilon_{\beta\delta\omega\epsilon}\bigg(\frac{k_{1\mu}k_{2\omega}}
{D_{\rho}(q+k_2)}+\frac{k_{2\mu}k_{1\omega}}{D_{\rho}(q+k_1)}\bigg).
\label{amp_back}
\end{equation}
\noindent Here $\delta $ is the additional phase (in this work we
treat it as a constant) taking into account $\rho\pi$ rescattering
effects \cite{rhophase}.

In the $K^+K^-$ loop model,  $\phi\to K^+K^-\to
\gamma(f_0+\sigma)$ \cite{achasov-89,achasov-97}, above the $K\bar
K$ threshold  the amplitude of the signal
$\phi\to\gamma(f_0+\sigma)\to\gamma\pi^0\pi^0$  is
\begin{equation}
M_{sig}=g(m)((\phi\epsilon)- \frac{(\phi q)(\epsilon
p)}{(pq)})T\left(K^+K^-\to\pi^0\pi^0\right )\times 16\pi
\label{f0signal}\,,
\end{equation}
where the $K^+K^-\to\pi^0\pi^0$ amplitude, taking into account the
mixing of $f_0$ and $\sigma$ mesons,
\begin{equation}
T\left(K^+K^-\to\pi^0\pi^0\right ) =
e^{i\delta_B}(\sum_{R,R'}\frac{g_{RK^+K^-}G_{RR'}^{-1}g_{R'\pi^0\pi^0}}{16\pi})\,,
\label{kktopipi}
\end{equation}
where $R,R'=f_0,\sigma$,
\begin{equation}
\delta_B =\delta^{\pi\pi}_B+\delta^{K\bar K}_B\,, \label{delph}
\end{equation}
where $\delta^{\pi\pi}_B$ and $\delta_B^{K\bar K}$ are phases of
the elastic background of the $\pi\pi$ and $K\bar K$ scattering,
respectively, see Refs. \cite{adsh-1980,z_phys,adsh-84}.

Note that the additional phase $\delta_B^{K\bar K}$ changes the
modulus of the amplitude at $m<2 m_{K}$. Let's define

\[P_K= \left\{\begin{array}{ll}
  e^{i\delta_B^{K\bar K}}\hspace{60 mm} m\geq 2
m_{K}\,;\\
 \mbox{analytical continuation of } e^{i\delta_B^{K\bar
 K}}\hspace{12 mm}
m<2 m_{K}\,.\hspace{53 mm}\addtocounter{equation}{1}
(\theequation)
 \label{Kphas}
\end{array}\right.\]

Note also that the phase $\delta_B^{\pi\pi}$ was defined as
$\delta_B$ in Refs. \cite{achasov-97,a0f0}.

The matrix of inverse propagators \cite{achasov-97} is

\[G_{RR'}\equiv G_{RR'}(m)=\left( \begin{array}{cc} D_{f_0}(m)&-\Pi_{f_0\sigma}(m)\\-\Pi_{f_0 \sigma}(m)&D_{\sigma}(m)\end{array}\right),\]

$$\Pi_{f_0 \sigma}(m)=\sum_{a,b} \frac{g_{\sigma ab}}{g_{f_0
ab}}\Pi^{ab}_{f_0}(m)+C_{f_0\sigma},$$

\noindent where the constant $C_{f_0\sigma}$ incorporates the
subtraction constant for the transition $f_0(980)\to(0^-0^-)\to
\sigma (600)$ and effectively takes into account contribution of
multi-particle intermediate states to $f_0\leftrightarrow\sigma$
transition, see Ref. \cite{achasov-97}. The inverse propagator of
the R scalar meson is presented in Refs.
\cite{adsh-79,adsh-80,adsh-1980,z_phys,adsh-84,our_prop,achasov-89,achasov-97,achasov-95}:

\begin{equation}
\label{propagator} D_R(m)=m_R^2-m^2+\sum_{ab}[Re
\Pi_R^{ab}(m_R^2)-\Pi_R^{ab}(m^2)],
\end{equation}
where $\sum_{ab}[Re \Pi_R^{ab}(m_R^2)-
\Pi_R^{ab}(m^2)]=Re\Pi_R(m_R^2)- \Pi_R(m^2)$ takes into account
the finite width corrections of the resonance which are the one
loop contribution to the self-energy of the $R$ resonance from the
two-particle intermediate  $ab$ states.

For pseudoscalar $ab$ mesons and $m_a\geq m_b,\ m\geq m_+$ one
has:

\begin{eqnarray}
\label{polarisator}
&&\Pi^{ab}_R(m^2)=\frac{g^2_{Rab}}{16\pi}\left[\frac{m_+m_-}{\pi
m^2}\ln \frac{m_b}{m_a}+\right.\nonumber\\
&&\left.+\rho_{ab}\left(i+\frac{1}{\pi}\ln\frac{\sqrt{m^2-m_-^2}-
\sqrt{m^2-m_+^2}}{\sqrt{m^2-m_-^2}+\sqrt{m^2-m_+^2}}\right)\right]
\end{eqnarray}
При $m_-\leq m<m_+$
\begin{eqnarray}
&&\Pi^{ab}_{R}(m^2)=\frac{g^2_{Rab}}{16\pi}\left[\frac{m_+m_-}{\pi
m^2}\ln \frac{m_b}{m_a}-|\rho_{ab}(m)|+\right.\nonumber\\
&&\left.+\frac{2}{\pi}|\rho_{ab}(m)
|\arctan\frac{\sqrt{m_+^2-m^2}}{\sqrt{m^2-m_-^2}}\right].
\end{eqnarray}
При $m<m_-$
\begin{eqnarray}
&&\Pi^{ab}_{R}(m^2)=\frac{g^2_{Rab}}{16\pi}\left[\frac{m_+m_-}{\pi
m^2}\ln \frac{m_b}{m_a}-\right.\nonumber\\
&&\left.-\frac{1}{\pi}\rho_{ab}(m)\ln\frac{\sqrt{m_+^2-m^2}-
\sqrt{m_-^2-m^2}}{\sqrt{m_+^2-m^2}+\sqrt{m_-^2-m^2}}\right].
\end{eqnarray}
\noindent и
\begin{equation}
\label{rho-ab}
\rho_{ab}(m)=\sqrt{(1-\frac{m_+^2}{m^2})(1-\frac{m_-^2}{m^2})}\,\,,\qquad
m_{\pm}=m_a\pm m_b
\end{equation}

The constants  $g_{Rab}$ are related to the width
\begin{equation}
\Gamma(R\to ab,m)=\frac{g_{Rab}^2}{16\pi m}\rho_{ab}(m).
\label{f0pipi}
\end{equation}

Note that we take into account intermediate states $\pi\pi,K\bar
K,\eta\eta,\eta '\eta,\eta'\eta'$ in the $f_0(980)$ and $\sigma
(600)$ propagators:

\begin{equation}
\Pi_{f_0}=\Pi_{f_0}^{\pi^+\pi^-}+\Pi_{f_0}^{\pi^0\pi^0}+\Pi_{f_0}^{K^+K^-}+
\Pi_{f_0}^{K^0\bar{K^0}}+\Pi_{f_0}^{\eta \eta}+\Pi_{f_0}^{\eta '
\eta}+\Pi_{f_0}^{\eta ' \eta '} ,
\end{equation}

\noindent and also for the $\sigma (600)$. We use
$g_{f_0K^0\bar{K^0}}=g_{f_0K^+K^-},
g_{f_0\pi^0\pi^0}=g_{f_0\pi^+\pi^-}/\sqrt{2}$, the same for the
$\sigma (600)$, too.

For other coupling constants we use the four-quark model
prediction \cite{achasov-89,achasov-84}:

$$g_{f_0\eta \eta}=-g_{f_0\eta '\eta '}=\frac{2\sqrt{2}}{3}\,
g_{f_0K^+K^-},\,\,g_{f_0\eta
'\eta}=-\frac{\sqrt{2}}{3}\,g_{f_0K^+K^-}\,;$$

$$g_{\sigma\eta \eta}=g_{\sigma\eta \eta '}=\frac{\sqrt{2}}{3}\,
g_{\sigma \pi^+\pi^-},\,\,g_{\sigma\eta '\eta
'}=\frac{1}{3\sqrt{2}}\,g_{\sigma \pi^+\pi^-}\,.$$

In the $K^+K^-$ loop model $g(m)$ has the following forms (see
Refs. \cite{achasov-89,achasov-01a,achasov-95,our_a0}).

\ \ \  For $m<2m_{K^+}$
\begin{eqnarray}
&&g(m)=\frac{e}{2(2\pi)^2}g_{\phi K^+K^-}\Biggl\{
1+\frac{1-\rho^2(m^2)}{\rho^2(m^2_{\phi})-\rho^2(m^2)}\times\nonumber\\
&&\Biggl[2|\rho(m^2)|\arctan\frac{1}{|\rho(m^2)|}
-\rho(m^2_{\phi})\lambda(m^2_{\phi})+i\pi\rho(m^2_{\phi})-\nonumber\\
&&-(1-\rho^2(m^2_{\phi}))\Biggl(\frac{1}{4}(\pi+
i\lambda(m^2_{\phi}))^2- \nonumber\\
&&-\Biggl(\arctan\frac{1}{|\rho(m^2)|}\Biggr)^2
\Biggr)\Biggr]\Biggr\},
\end{eqnarray}
where
\begin{equation}
\rho(m^2)=\sqrt{1-\frac{4m_{K^+}^2}{m^2}}\,\,;\qquad
\lambda(m^2)=\ln\frac{1+\rho(m^2)}{1-\rho(m^2)}\,\,;\qquad
\frac{e^2}{4\pi}=\alpha=\frac{1}{137}\,\,.
\end{equation}

 For $m\geq 2m_{K^+}$
\begin{eqnarray}
&&g(m)=\frac{e}{2(2\pi)^2}g_{\phi K^+K^-}\Biggl\{
1+\frac{1-\rho^2(m^2)}{\rho^2(m^2_{\phi})-\rho^2(m^2)}\times\nonumber\\
&&\times\Biggl[\rho(m^2)(\lambda(m^2)-i\pi)-
\rho(m^2_{\phi})(\lambda(m^2_{\phi})-i\pi)-\nonumber\\
&&\frac{1}{4}(1-\rho^2(m^2_{\phi}))
\Biggl((\pi+i\lambda(m^2_{\phi}))^2-
(\pi+i\lambda(m^2))^2\Biggr)\Biggr]\Biggr\}.
\end{eqnarray}

The mass spectrum of the reaction is
\begin{equation}
\frac{\Gamma(\phi\to\gamma\pi^0\pi^0)}{dm}=\frac{d\Gamma_{S}}{dm}+
\frac{d\Gamma_{back}(m)}{dm}+\frac{d\Gamma_{int}(m)}{dm},
\end{equation}
where the signal contribution $\phi\to S\gamma \to\pi^0\pi^0\gamma
$
 \begin{equation}
\frac{d\Gamma_{S}}{dm}=\frac{|P_K|^2
|g(m)|^2\sqrt{m^2-4m_{\pi}^2}(m_{\phi}^2-m^2)}
{3(4\pi)^3m_{\phi}^3}|\sum_{R,R'}g_{RK^+K^-}G_{RR'}^{-1}g_{R'\pi^0\pi^0}|^2.
\label{f0}
\end{equation}

The mass spectrum of the background process
$\phi\to\rho\pi^0\to\pi^0\pi^0\gamma$

\begin{equation}
\frac{d\Gamma_{back}(m)}{dm}=\frac{1}{2}\frac{(m_{\phi}^2-m^2)\sqrt{m^2-4m_{\pi}^2}}
{256\pi^3m_{\phi}^3} \int_{-1}^{1}dxA_{back}(m,x)\,,
\label{phonf0}
\end{equation}

where

\begin{eqnarray}
&&A_{back}(m,x)=\frac{1}{3}\sum|M_{back}|^2=\nonumber \\
&&=\frac{1}{24}g_{\phi\rho\pi}^2g_{\rho\pi\gamma}^2 \{
(m_{\pi}^8+2m^2m_{\pi}^4\tilde{m_{\rho}^2}-
4m_{\pi}^6\tilde{m_{\rho}^2}+2m^4\tilde{m_{\rho}^4}-\nonumber \\
&&4m^2m_{\pi}^2\tilde{m_{\rho}^4}+
6m_{\pi}^4\tilde{m_{\rho}^4}+2m^2\tilde{m_{\rho}^6}-4m_{\pi}^2\tilde{m_{\rho}^6}+
\tilde{m_{\rho}^8}-2m_{\pi}^6m_{\phi}^2-\nonumber \\
&&2m^2m_{\pi}^2\tilde{m_{\rho}^2}m_{\phi}^2+
2m_{\pi}^4\tilde{m_{\rho}^2}m_{\phi}^2-2m^2\tilde{m_{\rho}^4}m_{\phi}^2+
2m_{\pi}^2\tilde{m_{\rho}^4}m_{\phi}^2-2\tilde{m_{\rho}^6}m_{\phi}^2+\nonumber
\\ &&m_{\pi}^4m_{\phi}^4+ \tilde{m_{\rho}^4}m_{\phi}^4)
(\frac{1}{|D_{\rho}(\tilde{m}_{\rho})|^2}+
\frac{1}{|D_{\rho}(\tilde{m}_{\rho}^*)|^2})+(m_{\phi}^2-m^2)(m^2-\nonumber\\
&& 2m_{\pi}^2+2\tilde{m_{\rho}^2}-m_{\phi}^2)
(2m^2m_{\pi}^2+2m_{\pi}^2m_{\phi}^2-m^4)\frac{1}{|D_{\rho}(\tilde{m}_{\rho}^*)|^2}+
\nonumber \\
&&2Re(\frac{1}{D_{\rho}(\tilde{m}_{\rho})D^*_{\rho}(\tilde{m}_{\rho}^*)})
(m_{\pi}^8- m^6\tilde{m_{\rho}^2}+2m^4m_{\pi}^2\tilde{m_{\rho}^2}+
\nonumber \\
&&2m^2m_{\pi}^4\tilde{m_{\rho}^2}-4m_{\pi}^6\tilde{m_{\rho}^2}-
4m^2m_{\pi}^2\tilde{m_{\rho}^4}+6m_{\pi}^4\tilde{m_{\rho}^4}+\nonumber
\\ &&2m^2\tilde{m_{\rho}^6}-4m_{\pi}^2\tilde{m_{\rho}^6}+\tilde{m_{\rho}^8}+
m^2m_{\pi}^4m_{\phi}^2-2m_{\pi}^6m_{\phi}^2+
2m^4\tilde{m_{\rho}^2}m_{\phi}^2-\nonumber \\
&&4m^2m_{\pi}^2\tilde{m_{\rho}^2}m_{\phi}^2+
2m_{\pi}^4\tilde{m_{\rho}^2}m_{\phi}^2-m^2\tilde{m_{\rho}^4}m_{\phi}^2+
2m_{\pi}^2\tilde{m_{\rho}^4}m_{\phi}^2-2\tilde{m_{\rho}^6}m_{\phi}^2-\nonumber
\\ &&m_{\pi}^4m_{\phi}^4-m^2\tilde{m_{\rho}^2}m_{\phi}^4+
2m_{\pi}^2\tilde{m_{\rho}^2}m_{\phi}^4+\tilde{m_{\rho}^4}m_{\phi}^4)
\}, \label{aback}
\end{eqnarray}

\begin{eqnarray}
&&\tilde{m_{\rho}}^2=m_{\pi}^2+\frac{(m_{\phi}^2-m^2)}{2}
(1-x\sqrt{1-\frac{4m_{\pi}^2}{m^2}})\nonumber \\
&&\tilde{m_{\rho}}^{*2}=m^2_{\phi}+2m_{\pi}^2-m^2-\tilde{m_{\rho}}^2\,.
\end{eqnarray}

Note that in Ref. \cite{a0f0} there are typos, $m_\rho$ should be
replaced by $\tilde{m_\rho}$ everywhere in Eq. (18) of Ref.
\cite{a0f0}, as in Eq. (\ref{aback}) of the given paper, see also
Ref. \cite{bramon-2002}. Note also that all calculations in Ref.
\cite{a0f0} were done with the correct expression.

The interference between signal and background processes accounts
for

\begin{equation}
\frac{d\Gamma_{int}(m)}{dm}=\frac{1}{\sqrt{2}}\frac{\sqrt{m^2-4m_{\pi}^2}}
{256\pi^3m_{\phi}^3} \int_{-1}^{1}dxA_{int}(m,x)\,,
 \label{intf0}
\end{equation}
where

\begin{eqnarray}
&&A_{int}(m,x)=\frac{2}{3}(m_\phi^2-m^2)Re\sum M_f
M_{back}^*=\nonumber
\\
&&\frac{1}{3}Re\{P_K
e^{i\delta_B^{\pi\pi}}e^{i\delta}g(m)g_{\phi\rho\pi}g_{\rho\pi^0\gamma}
(\sum_{R,R'}g_{RK^+K^-}G_{RR'}^{-1}g_{R'\pi^0\pi^0})
(\frac{((\tilde{m_{\rho}^2}-m_{\pi}^2)^2m_{\phi}^2-(m_{\phi}^2-m^2)^2\tilde{m_{\rho}^2}
)} {D_{\rho}^*(\tilde{m_{\rho}})}+\nonumber\\
&&\frac{((\tilde{m_{\rho}^{*2}}-m_{\pi}^2)^2m_{\phi}^2-(m_{\phi}^2-m^2)^2
\tilde{m_{\rho}^{*2}} )}{D_{\rho}^*(\tilde{m_{\rho}}^*)})\}\,.
\label{f0int}
\end{eqnarray}

The factor $1/2$ in Eq. (\ref{phonf0}) and the factor $1/\sqrt{2}$
in Eq. (\ref{intf0}) take into account the identity of pions, the
same reason for definition
$g_{R\pi^0\pi^0}=g_{R\pi^+\pi^-}/\sqrt{2}$ in Eq. (\ref{f0}).

The S-wave amplitude $T^0_0$ of the $\pi\pi$ scattering with I=0
\cite{adsh-1980,z_phys,adsh-84,achasov-97} is

\begin{equation}
T^0_0=\frac{\eta^0_0
e^{2i\delta_0^0}-1}{2i\rho_{\pi\pi}(m)}=
\frac{e^{2i\delta_B^{\pi\pi}}-1}{2i\rho_{\pi\pi}(m)}+
e^{2i\delta_B^{\pi\pi}}\sum_{R,R'}\frac{g_{R\pi\pi}G_{RR'}^{-1}g_{R'\pi\pi}}{16\pi}\,.
\label{pipiamp}\end{equation}

Here $\eta^0_0\equiv \eta^0_0(m)$ is the inelasticity,
$\eta^0_0=1$ for $m\leq 2m_{K^+}$, and

\begin{equation}
\label{phas2}
\delta_0^0\equiv\delta_0^0(m)=\delta^{\pi\pi}_B(m)+\delta_{res}(m)\,,
\end{equation}

\noindent where $\delta_B^{\pi\pi}=\delta_B^{\pi\pi}(m)$
($\delta_B$ in Ref. \cite{achasov-97}) is the phase of the elastic
background (see Eq. \ref{delph}), and $\delta_{res}(m)$ is half of
the phase of

\begin{equation}
S_0^{0\ res}=\eta^0_0(m) e^{2i\delta_{res}(m)}=1+2i
\rho_{\pi\pi}(m)
\sum_{R,R'}\frac{g_{R\pi\pi}G_{RR'}^{-1}g_{R'\pi\pi}}{16\pi}\,,\,\,\,\eta^0_0=|S_0^{0\
res}|\,, \label{phR} \end{equation}

\noindent $g_{R\pi\pi}=\sqrt{3/2}\,g_{R\pi^+\pi^-}$. The chiral
shielding phase  $\delta_B^{\pi\pi}(m)$, motivated by the
$\sigma$-model Ref. \cite{annshgn-94}, is taken in the form
\begin{equation}\tan ( \delta_B^{\pi\pi})=-\frac{p_\pi}{m_\pi}
\bigg(b_0-b_1\frac{p_\pi ^2}{m_\pi^2}+ b_2\frac{p_\pi
^4}{m_\pi^4}\bigg)\frac{1}{1+(2p_\pi)^2
/\Lambda^2},\label{phB}\end{equation}

\noindent and

\begin{equation}
e^{2i\delta_B^{\pi\pi}}=\frac{1-i\frac{p_\pi}{m_\pi}
\bigg(b_0-b_1\frac{p_\pi ^2}{m_\pi^2}+ b_2\frac{p_\pi
^4}{m_\pi^4}\bigg)\frac{1}{1+(2p_\pi)^2
/\Lambda^2}}{1+i\frac{p_\pi}{m_\pi} \bigg(b_0-b_1\frac{p_\pi
^2}{m_\pi^2}+ b_2\frac{p_\pi
^4}{m_\pi^4}\bigg)\frac{1}{1+(2p_\pi)^2 /\Lambda^2}} \label{phB2}
\end{equation}

Here $2p_\pi=\sqrt{m^2-4m_{\pi}^2}$, and $(1+(2p_\pi)^2
/\Lambda^2)^{-1}$ is a cut-off factor. The phase $\delta_B^{K \bar
K}=\delta_B^{K \bar K}(m)$ is parameterized in the following way:

\begin{equation}
\tan \delta_B^{K \bar K}=f_K(m^2)\sqrt{m^2-4m^2_{K^+}}\equiv 2p_K
f_K(m^2)\label{phK}
\end{equation}
\noindent and

\begin{equation}
e^{2i\delta_B^{K\bar K}}=\frac{1+i2p_K f_K(m^2)}{1-i2p_K
f_K(m^2)}\label{phK2}
\end{equation}

Actually, $e^{2i\delta^{\pi\pi}_B(m)}$ has a pole at $m^2=m_0^2$,
$ 0<m^2_0<4m^2_\pi$ (details are below), which is compensated by
the zero in $e^{2i \delta_B^{K \bar K}(m)}$ to ensure a regular
$K\bar K\to\pi\pi$ amplitude and, consequently, the $\phi\to
K^+K^-\to\pi\pi\gamma$ amplitude at $ 0<m^2<4m_\pi^2$. This
requirement leads to
\begin{equation}
f_K(m_0^2)=\frac{1}{\sqrt{4m^2_{K^+}-m_0^2}}\approx
\frac{1}{2m_{K^+}}\,. \label{kcondition}
\end{equation}

The inverse propagator of the $\rho$ meson has the following
expression
\begin{equation}
D_{\rho}(m)=m_{\rho}^2-m^2-im^2\frac{g^2_{\rho\pi\pi}}{48\pi}
\bigg(1-\frac{4m_{\pi}^2}{m^2}\bigg)^{3/2}\,.
\end{equation}.

The coupling constants $g_{\phi K^+K^-}=4.376\pm 0.074$ and
$g_{\phi\rho\pi}=0.814\pm 0.018$ GeV$^{-1}$ are taken from the
most precise measurement Ref.\cite{sndphi}. Note that in Ref.
\cite{publ,a0f0} the value $g_{\phi K^+K^-}=4.59$ was obtained
using the \cite{pdg} data. To obtain the coupling constant
$g_{\rho\pi^0\gamma}$ we used the data of the experiments Ref.
\cite{dolinsky} and Ref. \cite{pigam} on the $\rho\to\pi^0\gamma$
decay and the expression

\begin{equation}
\Gamma(\rho\to\pi^0\gamma)=\frac{g_{\rho\pi^0\gamma}^2}{96\pi
m_{\rho}^3} (m_{\rho}^2-m_{\pi}^2)^3,
\end{equation}

\noindent the result $g_{\rho\pi^0\gamma}=0.26\pm 0.02$ GeV$^{-1}$
is the weighed average of these experiments.


\section{Data analysis}
\label{sr} \subsection{Restrictions}\label{restr}

To analyze the data, we construct a function to minimize:

\begin{equation}
\tilde{\chi}^2_{tot}=\chi^2_{sp}+\chi^2_{ph}.
\end{equation}

Here $\chi^2_{sp}$ is an usual $\chi^2$ function to fit
$\pi^0\pi^0$ spectrum in the $\phi\to\pi^0\pi^0\gamma$ decay, see
APPENDIX I for details, while $\chi^2_{ph}$ is the $\delta^0_0$
contribution. For the $\pi\pi$ scattering phase $\delta_0^0$ we
use the data \cite{hyams,estabrook,martin,srinivasan,rosselet}.


Some parameters are fixed by the requirement of the proper
analytical continuation of amplitudes. The phase factor
$e^{2i\delta_B^{\pi\pi}}$ has a singularity at $m_0^2$,
$0<m^2_0<4m^2_\pi$, when
\begin{equation}
1-\frac{4m_\pi^2-m_0^2}{\Lambda^2}-\frac{\sqrt{4m_\pi^2-m_0^2}}{2m_\pi}
\bigg(b_0+b_1\frac{4m_\pi^2-m_0^2}{(2m_\pi)^2}+
b_2\frac{(4m_\pi^2-m_0^2)^2}{(2m_\pi)^4}\bigg)=0\,,
\end{equation}

\noindent see Eq. (\ref{phB2}). Inasmuch as the amplitude
(\ref{pipiamp}) should have no poles at $0<m^2_0<4m^2_\pi$, Eq.
(\ref{phR}) should be equal to zero at the $m_0^2$ Ref.
\cite{nozeroes}. This condition fixes one free parameter. Another
free parameter may be removed by fixing the $\pi\pi$ scattering
length $a_0^0$
\begin{equation}
\sum_{R,R'}g_{R\pi\pi}G_{RR'}^{-1}g_{R'\pi\pi}\bigg|_{m=2m_{\pi^+}}-b_0=a_0^0\,.
\label{sclen}
\end{equation}
\noindent  In two variants of fitting the data (see below Fits 3,
4 in Table I) we take $a_0^0=0.22\ m_{\pi}^{-1}$ from the recent
calculation \cite{scat_len} based on the chiral perturbation
theory and Roy equations. This number excellently consists with
all other results, for example, with the BNL experimental value
$a^0_0=(0.228\pm 0.012)\ m_{\pi}^{-1}$ Ref. \cite{scatbnl}. In
other variants of the data analysis we treat $a_0^0$ as a free
parameter but all obtained values of $a_0^0$ are close to the BNL
one.

The Adler zeros is the separate question. The zero in the
amplitude of the $\pi\pi$ scattering appears automatically near
150 MeV in all obtained analyzes, see Tables I, II, while for the
amplitude Eq. (\ref{f0signal})  the Adler zero existence is a
rather strict constraint.

\subsection{Investigating the general scenario and around}

Analyzing data we imply a scenario motivated by the four-quark
model \cite{jaffe}, that is, the $\sigma$(600) coupling with the
$K\bar K$ channel $g_{\sigma K^+K^-}$ is suppressed relative to
one with the $\pi^+\pi^-$ channel $g_{\sigma \pi^+\pi^-}$, the
mass of the $\sigma$ meson is in the 500-700 MeV range.  In
addition, we have in mind the  Adler self-consistency conditions
for $T_0^0(\pi\pi\to\pi\pi)$ and $T(\phi\to\pi\pi\gamma)$ (i.e.,
in $T_0^0(K^+K^-\to\pi\pi)$) amplitudes near $\pi\pi$ threshold.
The general aim of this subsection is to demonstrate that the data
are in excellent agreement with this general scenario.

As for the $\pi\pi$ scattering amplitude, the Adler zero appears a
bit lower the $\pi\pi$ threshold automatically in all variants,
see Tables I, II and III.

The $\phi $ decay amplitude is  another matter.  The analysis
shows that the data prefer to have the zero in the amplitude of
the $\phi $ decay at the negative values of $m^2\sim 1$ GeV$^2$,
i.e., in the region of the left cut, see Table III. That is why we
require the $\phi$-decay amplitude  to have the Adler zero in the
interval $0<m^2<4m_{\pi}^2$, see Tables I and II.

Besides, the data favor negative $f_K(m^2)$ in the resonance
region, at $m>$700 MeV. A fit with the "effective" constant
$f_K(m^2)=\Delta_K$ gives $\Delta_K \approx -0.6/$GeV. But Eq.
(\ref{kcondition}) shows that near the $\pi\pi$ threshold the
function $f(m^2)$ should be about 1/GeV, i.e., positive. We choose
$f_K(m^2)$ in the form

\begin{equation}
f_K(m^2)=-\frac{\arctan (\frac{m^2-m_1^2}{m_2^2})}{\Lambda _K}
\label{fK}
\end{equation}
to get the desirable change of the sign at $m=m_1\approx 500-800$
MeV  in a rather simple way [only two new parameters are
introduced for Eq. (\ref{kcondition})]. Unfortunately, now we
haven't enough data and theory to determine $f_K(m^2)$ more
accurately.

The inelasticity $\eta^0_0(m)$ and the phase $\delta ^{\pi K}(m)$
of the amplitude $T(\pi\pi\to K\bar K)$ are essential in the fit
region, $2m_{K^+}< m< 1.1$ GeV. As for the inelasticity, the
experimental data of Ref. \cite{hyams} gives an evidence in favor
of low values of $\eta^0_0(m)$ near the $K\bar K$ threshold. The
situation with the experimental data on $\delta ^{\pi K}(m)$ is
controversial and experiments have large errors. We consider these
data as a guide, which main role is to fix the sign between signal
and background amplitudes (\ref{f0signal}) and (\ref{amp_back}),
and hold two points of the experiment \cite{pkphase}, see Fig. 7.

Providing all the conditions cited above, we set the Adler zero
position in $\pi\pi\to K\bar K $ to $m^2=m_\pi^2$ in  Fits 1-3 and
5-8,  listed in Tables I and II. All figures, Figs. 1-7,
correspond to  Fit 1 in Table I. Emphasize that  Fits 1, 3, 5, 7,
9  and 2,4, 6, 8, 10  correspond to the positive and negative
$g_{\sigma\pi^+\pi^-}/g_{f_0 K^+K^-}$ ratio, respectively.  The
sign of this ratio  reveals itself only in the $f_0 -\sigma$
mixing, which is reasonably small. That is why we obtain a kind of
symmetry  by slight change of the rest parameters, especially the
constant $C_{f_0\sigma}$. This consideration is supported by the
data, see Tables I, II, III and by the fact that the resonance
phase $\delta_{res} (m)$ reaches 90 and 270 degrees close to
$m_\sigma$ and $m_{f_0}$ respectively.

As seen from Tables, the obtained variants don't leave any doubt
that the data are in perfect agreement with the general scenario,
which we discuss.

A crucial aspect of the data description is a low inelasticity.
The key experimental point is $\eta^0_0 (m=1.01 \mbox{ GeV} )=
0.41\pm 0.14$, see Fig. 4. To demonstrate the possibility of the
low inelasticity in our model we list two variants in Table I with
the fixed inelasticities: Fit 3 with $\eta^0_0 (m=1.01 \mbox{ GeV}
)=0.39$ and Fit 4 with $\eta^0_0 (m=1.01 \mbox{ GeV} )=0.41$ with
the different signs of $g_{\sigma\pi^+\pi^-}/g_{f_0 K^+K^-}$.

In Fit 4 we set the Adler zero position in $\pi\pi\to K\bar K $ to
$m^2=m_\pi^2/2$ for variety. Note that varying the Adler zero
position in $\pi\pi\to K\bar K $ amplitude at $0<m^2<4m_{\pi}^2$
doesn't give noticeable effect.


The excellent description of the data gives the possibility to
study different physical situations and to find crucial points for
further investigations.

In particular, in Fits 5,6, Table II, we have the picture close to
the naive four-quark model \cite{jaffe}: $f_0(980)$ weakly couples
to the $\pi\pi$ channel, the $\sigma (600)$ practically does not
couple  to the $K\bar K$ channel, and $g_{\sigma\pi^+
\pi-}^2\approx 2 g_{f_0 K^+ K^-}^2$, as predicted by the naive
four-quark model, see Ref. \cite{jaffe,mistake}.

The data may be perfectly described with low $m_\sigma\approx 400$
MeV also, see Fits 7 and 8 in Table II. Remind that low mass of
$\sigma (600) $ is interesting for the sigma term problem.

Removing the requirement of the Adler zero in the $\pi\pi\to K
\bar K$ amplitude at $0<m^2<4m^2_\pi$ leads to a small improvement
of $\chi^2$ and inessential changes in the parameters. As for
Adler zero, it goes to $m^2\approx -(1$ GeV$)^2$, see Fits 9 and
10 in Table 3.

\begin{center}
\begin{tabular}{|c|c|c|c|c|}
\multicolumn{3}{c}{Table I. Results of the analysis, Fits 1-4.} \\
\hline

Fit & 1 & 2 & 3 & 4 \\ \hline $m_{f_0}$, MeV & $984.1$ & $985.2$ &
$984.8$ & $987.6$  \\ \hline

$g_{f_0K^+K^-}$, GeV  & $4.3$ & $4.2$ & $5.1$ & $4.7$   \\ \hline

$\frac{g_{f_0K^+K^-}^2}{4\pi}$, GeV$^2$  & 1.44 & 1.39 & 2.09 &
1.79
\\ \hline

$g_{f_0 \pi^+\pi^-}$, GeV  & $-1.8$ & $-2.0$ & $-1.9$ & $-2.1$  \\
\hline

$\frac{g_{f_0\pi^+ \pi^-}^2}{4\pi}$, GeV$^2$ & $0.25$ & $0.32$ &
$0.28$ & $0.36$   \\ \hline

 $m_{\sigma}$, MeV & $461.9$ & $485.0$ & $472.0$ & $542.6$   \\ \hline

$g_{\sigma\pi^+ \pi^-}$, GeV & $2.4$ & $-2.2$ & $2.5$ & $-2.5$
\\ \hline

$\frac{g_{\sigma\pi^+ \pi^-}^2}{4\pi}$, GeV$^2$ & $0.44$ & $0.38$
& $0.50$ & $0.49$   \\ \hline

$\Gamma _\sigma $, MeV & $286.0$ & $240.2$ & $319.7$ & $289.9$
\\ \hline

$g_{\sigma K^+K^-}$, GeV & $0.55$ & $-0.93$ & $0.43$ & $-1.1$
\\ \hline

$\frac{g_{\sigma K^+K^-}^2}{4\pi}$, GeV$^2$ & $0.024$ & $0.07$ &
$0.015$ & $0.10$   \\ \hline

$C$, GeV$^2$ & $0.047$ & $0.12$ & $-0.008$ & $0.13$  \\ \hline

$\delta $, $^{\circ}$ & $-11.4$& $-11.5$ & $-24.7$ & $-12.0$   \\
\hline

$b_0$ & $4.9$& 4.8 & 5.3 & $5.2$   \\ \hline

$b_1$ & $1.1$ & 1.1 & 0.90 & $0.55$  \\ \hline

$b_2$ & $1.36$ & 1.32  & 1.18 & $0.50$ \\ \hline

$\Lambda$, MeV & 172.2 & 172.8 & 160.0 & 150.3   \\ \hline

$m_1$, MeV & 765.4 & 766.4 & 795.6 & 784.1  \\ \hline

$m_2$, MeV & 368.9 & 368.6 & 375.7 & 362.2  \\ \hline

$\Lambda _K$, GeV & 1.24 & 1.24 & 1.25 & 1.26  \\ \hline

$a^0_0,\ m_\pi^{-1}$ & 0.209 & 0.209 & 0.22 & 0.22 \\ \hline

Adler zero in $\pi\pi\to\pi\pi $ & ($178$ MeV)$^2$ & ($207$
MeV)$^2$ & ($199$ MeV)$^2$ & ($185$ MeV)$^2$  \\ \hline

Adler zero in $\pi\pi\to K\bar K $ & $m_\pi^2$ & $m_\pi^2$ &
$m_\pi^2$ & $m_\pi^2/2$ \\ \hline

$\eta^0_0$($1010$ MeV) & $0.48$ & $0.48$ & $0.39$ & $0.41$  \\
\hline

$\chi^2_{tot}/n.d.f.$ (CL) & $40.0/48$ (79\%) & $40.0/48$ (79\%) &
$44.2/49$ (67\%) & $45.8/49$ (60\%)
\\ \hline

$\chi^2_{sp}$ ($18$ points) & $11.7$ & $11.6$ & $10.2$ & $11.8$
\\ \hline
\end{tabular}
\end{center}

\begin{center}
\begin{tabular}{|c|c|c|c|c|}
\multicolumn{3}{c}{Table II. Results of the analysis, Fits 5-8.}
\\ \hline

Fit & 5 & 6 & 7 & 8 \\ \hline $m_{f_0}$, MeV & $987.1$  & $984.2$
& $982.1$ & $982.1$ \\ \hline

$g_{f_0K^+K^-}$, GeV   & $2.9$ & $2.8$ & $4.2$ & $4.2$ \\ \hline

$\frac{g_{f_0K^+K^-}^2}{4\pi}$, GeV$^2$   & $0.67$ & $0.62$ &
$1.44$ & $1.42$ \\ \hline

$g_{f_0 \pi^+\pi^-}$, GeV   & $-0.9$ & $-0.8$ & $-1.7$ & $-1.7$ \\
\hline

$\frac{g_{f_0\pi^+ \pi^-}^2}{4\pi}$, GeV$^2$  & $0.07$ & $0.05$ &
$0.23$ & $0.23$ \\ \hline

$m_{\sigma}$, MeV  & $709.0$ & $692.5$ & $400$ & $415$ \\ \hline

$g_{\sigma\pi^+ \pi^-}$, GeV  & $3.6$ & $-3.3$ & $2.1$ & $-2.2$
\\ \hline

$\frac{g_{\sigma\pi^+ \pi^-}^2}{4\pi}$, GeV$^2$  & $1.01$ & $0.89$
& $0.36$ & $0.39$ \\ \hline

$\Gamma _\sigma $, MeV  & $492.5$ & $442.4$ & $241.9$ & $259.6$ \\
\hline

$g_{\sigma K^+K^-}$, GeV & $0.13$ & $-0.035$ & $0.38$ & $-0.37$
\\ \hline

$\frac{g_{\sigma K^+K^-}^2}{4\pi}$, GeV$^2$  & $0.001$ & $\approx
0$ & $0.01$ & $0.01$ \\ \hline

$C$, GeV$^2$ & $-0.01$ & $-0.02$ & $-0.015$ & $0.015$\\ \hline

$\delta $, $^{\circ}$  & $-9.8$ & $-6.4$ & $-28.0$ & $-20.5$ \\
\hline

$b_0$  & 4.6 & $3.0$ & $5.6$ & $5.4$ \\ \hline

$b_1$  & 1.2 & $0.86$ & $6.8$ & $3.7$ \\ \hline

$b_2$  & 0.18 & $0.16$ & $9.0$ & $5.0$ \\ \hline

$\Lambda$, MeV  & 149.4 & $181.0$ & $247.4$ & $200.4$ \\ \hline

$m_1$, MeV  & $577.5$ & $585.4$ & $753.4$ & $757.0$ \\ \hline

$m_2$, MeV  & $581.7$ & $680.4$ & $361.5$ & $362.9$ \\ \hline

$\Lambda _K$, GeV  & $0.62$ & $0.50$ & $1.24$ & $1.24$ \\ \hline

$a^0_0,\ m_\pi^{-1}$  & $0.213$ & $0.207$ & $0.215$ & $0.209$ \\
\hline

Adler zero in $\pi\pi\to\pi\pi $ & ($190$ MeV)$^2$ & ($194$
MeV)$^2$ & ($232$ MeV)$^2$ & ($226$ MeV)$^2$ \\ \hline

Adler zero in $\pi\pi\to K\bar K $ & $m_\pi^2$ & $m_\pi^2$ &
$m_\pi^2$ & $m_\pi^2$ \\ \hline

$\eta^0_0$($1010$ MeV)  & $0.56$ & $0.57$ & $0.50$ & $0.49$ \\
\hline

$\chi^2_{tot}/n.d.f.$ (CL)  & $48.1/48$ (47\%) & $47.1/48$ (51\%)
& $54.4/49$ (28\%) & $43.9/49$ (68\%) \\ \hline

$\chi^2_{sp}$ ($18$ points)  & $17.8$  & $15.6$ & $15.6$ &
$13.4$\\ \hline
\end{tabular}
\end{center}

\begin{center}
\begin{tabular}{|c|c|c|}
\multicolumn{3}{c}{Table III. Results of the analysis, Fits 9,10.}
\\ \hline

fit & 9 & 10 \\ \hline $m_{f_0}$, MeV &$983.2$ & $987.1$ \\ \hline

$g_{f_0K^+K^-}$, GeV  & $4.0$ & $3.7$ \\ \hline

$\frac{g_{f_0K^+K^-}^2}{4\pi}$, GeV$^2$  & $1.25$ & $1.06$
\\ \hline

$g_{f_0 \pi^+\pi^-}$, GeV  & $-1.3$ & $-1.9$ \\ \hline

$\frac{g_{f_0\pi^+ \pi^-}^2}{4\pi}$, GeV$^2$ & $0.15$ & $0.29$ \\
\hline

$m_{\sigma}$, MeV & $528.6$ & $566.3$ \\ \hline

$g_{\sigma\pi^+ \pi^-}$, GeV & $2.8$ &  $-2.4$
\\ \hline

$\frac{g_{\sigma\pi^+ \pi^-}^2}{4\pi}$, GeV$^2$ & $0.61$ &  $0.46$
\\ \hline

$\Gamma _\sigma $, MeV & $365.8$ & $263.8$ \\ \hline

$g_{\sigma K^+K^-}$, GeV & $1.1$ & $-1.8$ \\ \hline

$\frac{g_{\sigma K^+K^-}^2}{4\pi}$, GeV$^2$ & $0.09$ & $0.25$
\\ \hline

$C$, GeV$^2$ & $0.01$ & $0.14$\\ \hline

$\delta $, $^{\circ}$ & $-38.8$ & -37.2 \\ \hline

$b_0$ & $5.2$ & 5.22 \\ \hline

$b_1$ & $0.48$ & 0.43\\ \hline

$b_2$ & $0.43$ & 0.46\\ \hline

$\Lambda$, MeV & $149.0$ & 149.5 \\ \hline

$m_1$, MeV & $803.0$ & 801.3\\ \hline

$m_2$, MeV & $328.9$ & 330.6\\ \hline

$\Lambda _K$, GeV & $1.31$ & 1.31\\ \hline

$a^0_0,\ m_\pi^{-1}$ & $0.228$ & $0.229$\\ \hline

Adler zero in $\pi\pi\to\pi\pi $ & ($179$ MeV)$^2$ & ($179$
MeV)$^2$\\ \hline

Adler zero in $\pi\pi\to K\bar K $ & $-(1$ GeV$)^2$ & $-(0.9$
GeV$)^2$
\\ \hline

$\eta^0_0$($1010$ MeV) & $0.54$ & $0.54$\\ \hline

$\chi^2_{tot}/n.d.f.$ (CL) & $38.5/47$ (81\%) & $38.6/47$ (80\%).
\\ \hline

$\chi^2_{sp}$ ($18$ points) & $10.4$ & $10.5$\\ \hline
\end{tabular}
\end{center}

\begin{figure} \centerline{
\epsfxsize=12 cm \epsfysize=8cm \epsfbox{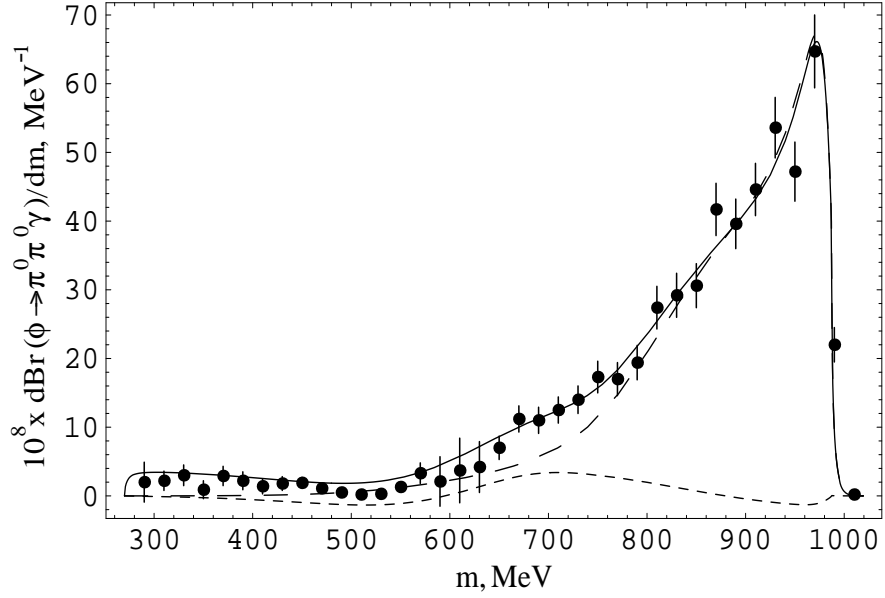}}
 \caption{The $\pi^0\pi^0$ spectrum, theoretical curve (solid line) and the KLOE data (points). The signal contribution and the interference term are
shown with the dashed line and the dotted line. All figures are
for Fit 1 in Table I} \label{fig2}
\end{figure}

\begin{figure} \centerline{
\epsfxsize=12 cm \epsfysize=8 cm \epsfbox{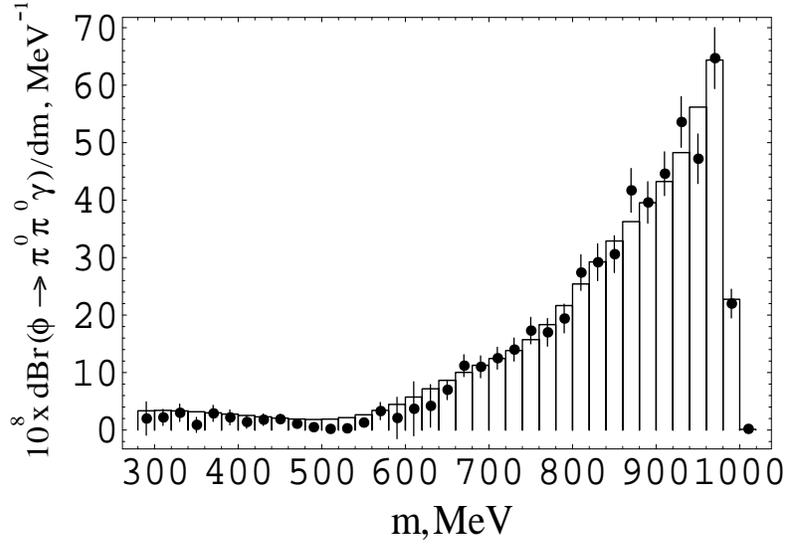}}
 \caption {The comparison of the Fit 1 and the KLOE data.
Histograms show Fit 1 curve averaged around each bin (see Appendix
I)} \label{fig1}
\end{figure}

\begin{figure} \centerline{
\epsfxsize=12 cm \epsfysize=8cm \epsfbox{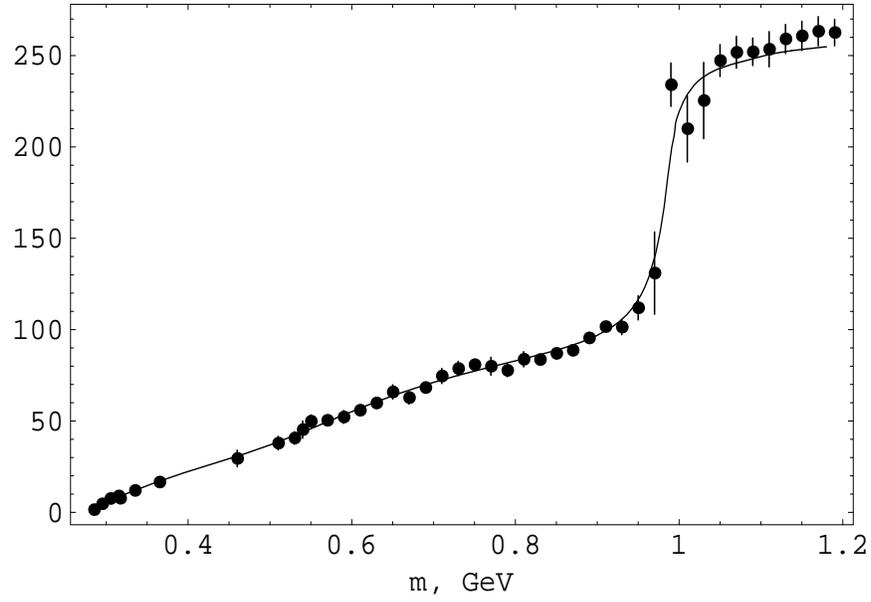}}
 \caption{The phase $\delta_0^0$ of the $\pi\pi$ scattering (degrees)} \label{fig3}
\end{figure}

\begin{figure} \centerline{
\epsfxsize=12 cm \epsfysize=8cm \epsfbox{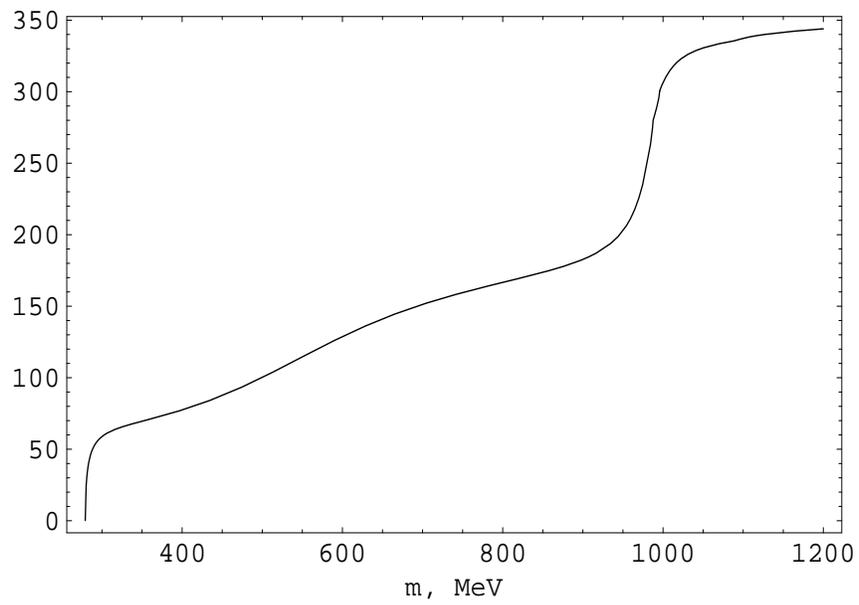}}
 \caption{The resonant phase $\delta_{res}$} \label{fig32}
\end{figure}

\begin{figure} \centerline{
\epsfxsize=12 cm \epsfysize=8cm \epsfbox{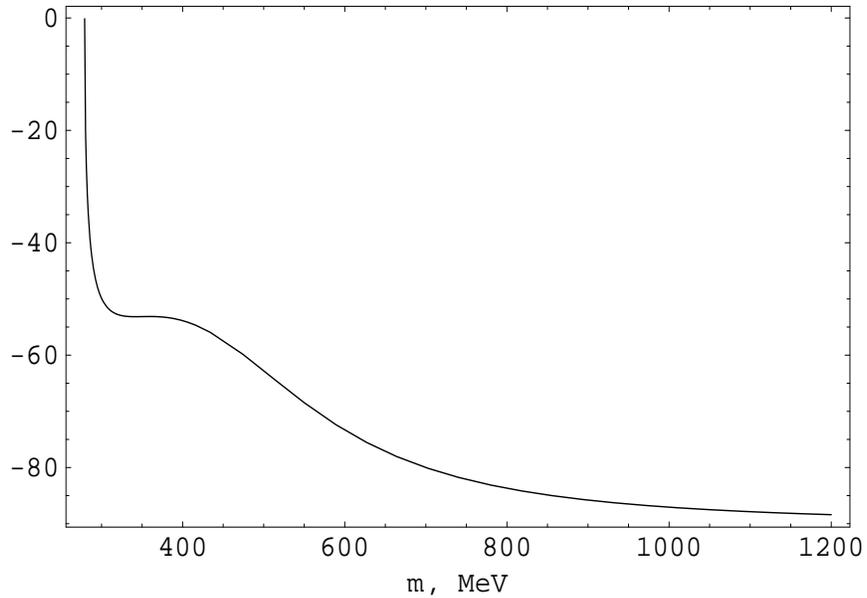}}
 \caption{The background phase $\delta^{\pi\pi}_{B}$} \label{fig34}
\end{figure}

\begin{figure} \centerline{
\epsfxsize=12 cm \epsfysize=8cm \epsfbox{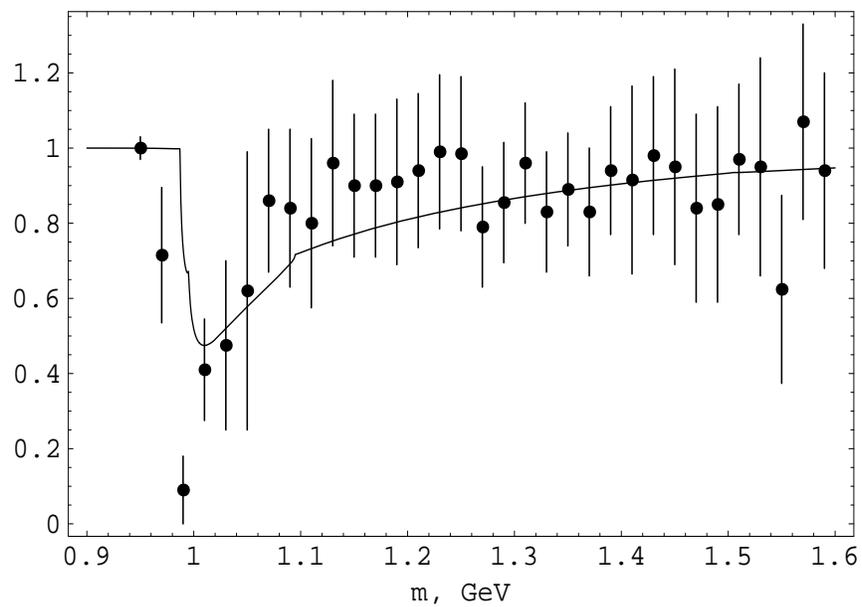}}
 \caption{The inelasticity $\eta^0_0$} \label{fig4}
\end{figure}

\begin{figure} \centerline{
\epsfxsize=12 cm \epsfysize=8cm \epsfbox{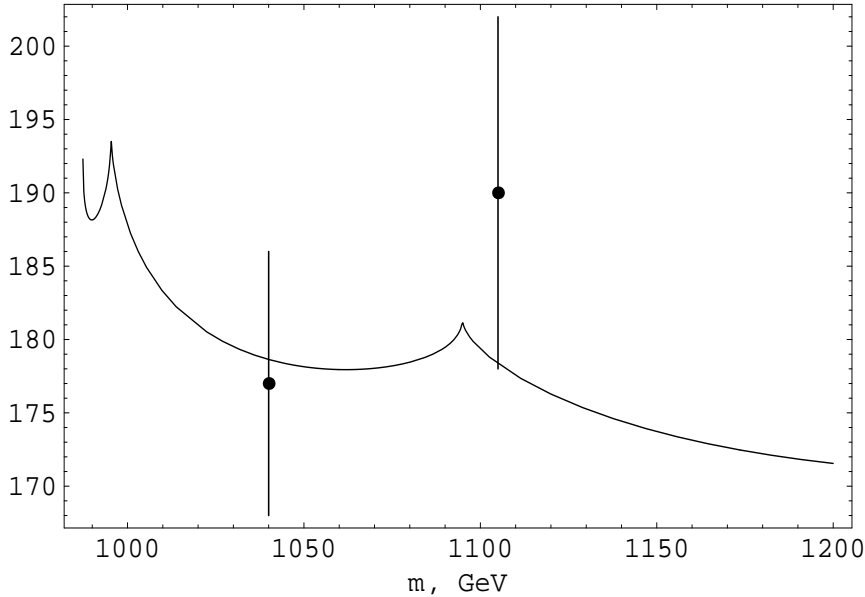}}
 \caption{The phase $\delta^{\pi K}$ of the $\pi\pi\to K\bar K $ scattering} \label{fig6}
\end{figure}

\section{Conclusion}
\label{sc}

So   the experimental data on the $\phi\to\pi^0\pi^0\gamma$ decay,
 the $\pi\pi$ scattering and the $\pi\pi\to K\bar K$ reaction
up to 1.1 GeV are
 perfectly described in
the model of the coupled $\sigma(600)$ and $f_0(980)$ resonances
under the general scenario (unitarity, analiticity of the
amplitudes, the Adler zeroes, the chiral shielding of
$\sigma(600)$,  $a_0^0\approx 0.22\ m_\pi^{-1}$).

To reduce (if not avoid) an effect of heavier isosinglet scalars
we restrict ourselves to the analysis of the mass region $m<1.1$
GeV$^2$, where, as one may expect, an effect of heavier scalars
could not be essential. As to mixing light and heavier isosinglet
scalars, this question could not be resolved once and for all at
present, in particular, because their properties are not
well-established up to now. A  preliminary consideration was
carried out in Ref. \cite{joe}, where, in particular, it was shown
that the mixing could affect  the mass difference of the isospinor
and isovector.

Of course, analyticity considerations are essential for our
analysis. It's well-known that an  analytical function can be
restored if its values at an interval are known. One can hope that
a description with good analytical characteristics is  close to
the exact one. So, one can consider the automatic formation of the
Adler zero in the $\pi\pi\to\pi\pi$ amplitude  at $0<m^2<4m^2_\pi$
as a wink  that we follow the right path. As for the $\pi\pi\to
K\bar K$ amplitude, we have to put the  Adler zero at
$0<m^2<4m^2_\pi$ as a constraint. To all appearance, it is a
prompt that this process is more complicated.

The weak coupling of $\sigma (600)$ with the $K\bar K$ channel and
$f_0(980)$ with the $\pi\pi$ channel practically in all variants
agrees qualitatively with the four-quark model \cite{jaffe}.
Certainly, there is also a suppression of the coupling of $\sigma
(600)$ with the $K\bar K$ channel and a strong suppression of the
coupling $f_0(980)$ with the $\pi\pi$ channel in the $q\bar q$
model: $\sigma(600) = (u\bar u + d\bar d)/\sqrt{2}$ and
$f_0(980)=s\bar s$. But the $f_0(980)$ and $a_0(980)$ mass
degeneracy cannot be explained in the naive two-quark model in
this case because $a_0(980) = (u\bar u - d\bar d)/\sqrt{2}$. In
addition, the photon spectra in $\phi\to\gamma
f_0(980)\to\gamma\pi^0\pi^0$ and $\phi\to\gamma
a_0(980)\to\gamma\pi^0\eta$ cannot be explained in this case also
\cite{achasov-2003}. Emphasize once more that Fits 5 and 6 in
Table II are especially interesting from the four-quark model
standpoint: $f_0(980)$ weakly couples to the $\pi\pi$ channel,
$\sigma (600)$ practically does not couple to the $K\bar K$
channel, and $g_{\sigma\pi^+ \pi-}^2\approx 2 g_{f_0 K^+ K^-}^2$.
 The practical absence of the $\sigma (600)$ coupling  with the
$K\bar K$ channel is excluded in the $q\bar q$ model, while it is
the characteristic of the lightest isoscalar scalar primary states
in the four-quark model \cite{jaffe}.

Note that in all variants $g_{f_0 K^+K^-}^2/4\pi$ has the same
order as $g_{a_0 K^+K^-}^2/4\pi$ obtained in Ref. \cite{our_a0}
which agrees reasonably with the four-quark model.

Our investigation   confirms  in full the $K^+K^-$ loop mechanism
of the $f_0(980)$ production which means the radiative four-quark
transition from $\phi(1020)$ to $f_0(980)$ and testifies to the
four-quark nature of $f_0(980)$ \cite{achasov-2003}.

The elucidation of the situation, a contraction of the possible
variants or even the selection of the unique variant, requires
considerable efforts. The new precise experiment on $\pi\pi\to
K\bar K$  would give  the crucial information about the
inelasticity $\eta^0_0$   and about the phase $\delta_B^{K\bar
K}(m)$ near the $K\bar K$ threshold.  The forthcoming precise
experiment in KLOE on the $\phi\to\pi^0\pi^0\gamma $ decay will
also help to judge about this phase in an indirect way. The
precise measurement of the inelasticity $\eta^0_0$ near 1 GeV in
$\pi\pi\to\pi\pi $ would be also very important.
 We hope also to
find  answers for some troubles in  consideration of  heavy
quarkonia decays, which are our next aim.  The new precise
experiment on $\gamma\gamma\to\pi\pi$ up to 1 GeV is urgent for an
understanding of the mechanism of the $\sigma(600)$ production and
hence for an understanding of its nature.
\\[1pc]

{\bf APPENDIX I: $\chi ^2$ FUNCTION FOR THE DATA ON THE
$\phi\to\pi^0\pi^0\gamma$ DECAY}\\[1pc]

In the experiment the whole mass region ($2m_{\pi^0},m_{\phi}$) is
divided into some number of bins. Experimenters measure the
average value ${\bar{B}_i}$ ("i" is the number of bin) of
$dBr(\phi\to\pi^0\pi^0\gamma)/dm$ around each i-th bin:

$$ \bar{B}_i=\frac{1}{m_{i+1}-m_i }\int ^{m_{i+1}}_{m_i}
dBr(\phi\to\pi^0\pi^0\gamma)/dm,$$

In this case one should define $\chi^2$ function as:

$$ \chi^2_{sp}=\sum_i
\frac{(\bar{B}_i^{th}-\bar{B}_i^{exp})^2}{\sigma _i^2},$$

\noindent where $\bar{B}_i^{exp}$ are the experimental results,
$\sigma _i$ are the experimental errors, and

$$ \bar{B}_i^{th}=\frac{1}{m_{i+1}-m_i }\int ^{m_{i+1}}_{m_i}
dBr^{th}(\phi\to\pi^0\pi^0\gamma)/dm $$

($dBr^{th}(\phi\to\pi^0\pi^0\gamma)/dm$ is the theoretical
curve).\\[1mm]

\begin{figure}
\centerline{ \epsfxsize=12 cm \epsfysize=8cm
\epsfbox{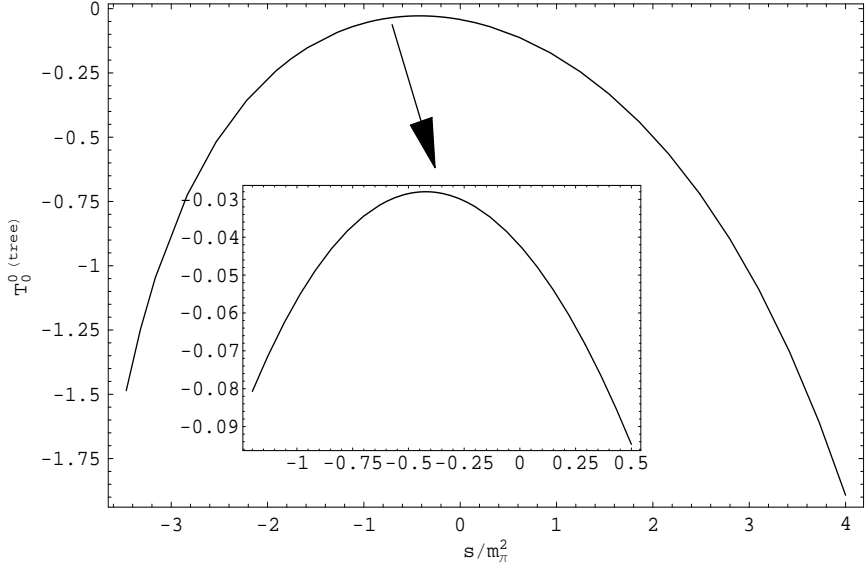}}
 \caption{ Plot of the amplitude $T^{0(tree)}_0$(m), for $m_\sigma $=400 MeV.}
\label{fig5}
\end{figure}
\vspace*{1pc}

{\bf APPENDIX II: About Adler zeroes}\\[1pc]

The well-known Adler condition Ref. \cite{adler-65b} means that
the amplitude $T^I(\pi\pi\to\pi\pi )\equiv A^I(s,t,u)$ with
isospin I satisfies

$$A^I(m_{\pi}^2,m_{\pi}^2,m_{\pi}^2)=0\,.$$

Though it doesn't mean, in general, that partial amplitudes
$T^I_l(\pi\pi\to\pi\pi )\equiv A^I_l (s,t,u)$ must have zeroes. To
illustrate this idea let's consider an example. In the linear
sigma model the amplitudes $A^I(s,t,u)$, calculated in the first
order in tree-level approximation, satisfy the Adler condition for
all values of $m_\sigma $, while the amplitude Ref.
\cite{basdevant}

$$T^{0(tree)}_0=\frac{m_\pi^2-m_\sigma^2}{F_\pi^2}\bigg[
5-3\frac{m_\sigma^2-m_\pi^2}{m_\sigma^2-s}-2\frac{m_\sigma^2-m_\pi^2}{s-4m_\pi^2}\ln\bigg(
1+\frac{s-4m_\pi^2}{m_\sigma^2} \bigg) \bigg] $$

\noindent has no zeroes for small $m_\sigma$. For example, for
$m_\sigma=$400 MeV the amplitude $T^{0(tree)}_0$ has no zeroes up
to $s=4m_\pi^2-m_\sigma^2\approx -4m_\pi^2$, where this amplitude
has a cut, see Fig. 8. Note that for $m_\sigma =500$ MeV
$T^{0(tree)}_0$ has a zero near $s=m_\pi^2/4$.

Hence a criticism and even a denial of some works basing on the
absence of the Adler zeroes is, in general, incorrect. However,
establishing  the Adler zero near the threshold is very convenient
because it guarantees weakness of $\pi\pi$ interactions near the
threshold, what is the main physical consequence of the Adler
condition of self-consistency. We can see it in the example
mentioned above: though the amplitude does not reach zero, its
absolute value is small near the threshold. That's why we believe
that both $T^0_0$ (Eq. (\ref{pipiamp})) and $M_{sig}$ (Eq.
(\ref{f0signal})) have the Adler zeroes not far from the
$s=4m_\pi^2$ threshold.

\section{Acknowledgements}
We thank S. Giovannella and S. Miscetti (KLOE Collaboration) very
much for providing the useful information, discussions and kind
contacts. This work was supported in part by the Presidential
Grant No. 2339.2003.2 for support of Leading Scientific Schools.
A.V.K. thanks very much the Dynasty Foundation and ICFPM for
support.


\begin{thebibliography}{99}
\bibitem{approxim}
A few approximate non-perturbative solutions in the frame of the
simplest linear $\sigma$-model are found in Ref.
\cite{annshgn-94}.
\bibitem{nozeroes}
Note that $e^{2i\delta_B^{\pi\pi}}$ also should have no (simple)
zeros, otherwise there are cuts in the amplitude Eq.
(\ref{f0signal}). Fortunately, in all obtained variants there are
no such zeros.
\end{thebibliography}
\end{document}